\documentclass{elsart}
\usepackage{epsfig}
\usepackage{rotating}
\begin{document}
\runauthor{Marco Battaglia}
\begin{frontmatter}
\title{CMOS Pixel Sensor Response to Low Energy Electrons
in Transmission Electron Microscopy}
\author[UCB,LBNL]{Marco Battaglia\corauthref{cor}},
\corauth[cor]{Corresponding author, Address: Lawrence 
Berkeley National Laboratory, 
Berkeley, CA 94720, USA, Telephone: +1 510 486 7029.} 
\ead{MBattaglia@lbl.gov}
\author[LBNL]{Devis Contarato},
\author[LBNL]{Peter Denes},
\author[LBNL]{Dionisio Doering},
\author[LBNL]{Velimir Radmilovic}
\address[UCB]{Department of Physics, University of California at 
Berkeley, CA 94720, USA}
\address[LBNL]{Lawrence Berkeley National Laboratory, 
Berkeley, CA 94720, USA}
\begin{abstract}
This letter presents the results of a study of the response of a test CMOS 
sensor with a radiation tolerant pixel cell design to 80~keV and 100~keV 
electrons. The point spread function is measured to be (13.0 $\pm$ 1.7)~$\mu$m 
at 100~keV and (12.1 $\pm$ 1.6)~$\mu$m at 80~keV, for 
20~$\mu$m pixels. Results agree well with values predicted by a Geant-4 
and dedicated charge collection simulation.
 
\end{abstract}
\end{frontmatter}

\typeout{SET RUN AUTHOR to \@runauthor}

%\linenumbers

\section{Introduction}

\vspace{0.5cm}

Monolithic CMOS pixel sensors open new perspectives in fast nano-imaging 
through single electron direct detection in transmission electron microscopy 
(TEM). High-voltage electron microscopy, developed and used for high resolution 
imaging in the late 1970s~\cite{nature} 
produced advances in spatial resolution, but was abandoned due to the severe 
displacement damage of the sample. As the displacement damage threshold is 
proportional to $\sqrt{E}$, there is now much interest in TEM of organic 
samples with energies of  80-100~keV, where recent advances in electron optics 
ensure deep sub-angstrom spatial resolution~\cite{science}. 
For example, the maximum energy transferred by an 80~keV electron to a carbon atom is 
15.6~eV, which is below the threshold for knock-on damage. This makes low energy TEM 
necessary for atomic resolution studies of samples such as single atomic layers of 
carbon in graphene or carbon nanotubes~\cite{Meyer} and in biology. 
There are two main issues to be considered for imaging 
with low energy electrons. The first is the large fluctuations in the energy 
deposition. The second is the degradation of the point spread 
function (PSF) due to the $1/E$ increase of the electron multiple 
scattering in the detector. 

In an earlier paper~\cite{Battaglia:2008yt}, we presented the design of a 
radiation tolerant CMOS pixel cell and investigated the response of 10~$\mu$m 
and 20~$\mu$m pixels to electrons in the energy range 120~keV up to 300~keV for 
TEM. In this letter we extend that study to lower energies, by investigating 
the response to 80~keV and 100~keV electrons. 

\section{Simulation}

We perform a detailed simulation of the charge deposition and signal formation 
in the CMOS pixel sensor based on the {\tt Geant-4} program~\cite{Agostinelli:2002hh}
using the low energy electromagnetic physics models~\cite{Chauvie:2001fh}.
The CMOS sensor is modelled according to the detailed geometric structure of oxide, 
metal interconnect and silicon layers, as specified by the foundry.
Electrons are incident perpendicular to the detector plane and tracked through 
the sensor. For each interaction within the epitaxial layer, the ionisation point 
position and the amount of energy released are recorded.

Charge collection in the sensor is simulated with {\tt PixelSim}, a dedicated 
digitisation module~\cite{Battaglia:2007eu}, developed in the {\tt Marlin} 
C++ reconstruction framework~\cite{Gaede:2006pj}, originally deployed for the 
International Linear Collider particle physics project. The processor starts 
from the ionisation points generated along the particle trajectory by {\tt Geant-4} 
and stored in {\tt lcio} format~\cite{Gaede:2003ip}. The {\tt PixelSim} simulation 
models diffusion of charge carriers from their production point in the epitaxial layer 
to the collection diode. This provides us with full simulation of the response of each 
individual pixels in the detector matrix, including sensor geometry and electronics 
noise effects, which can be processed through the same analysis chain as the 
experimental data. 
The simulation has a single free parameter, the diffusion parameter 
$\sigma_{{\mathrm{diff}}}$, used to determine the width of the charge carrier cloud. 
Its value is extracted  from data by a $\chi^2$ fit to the pixel multiplicity in the 
clusters of 1.5~GeV electrons since, at this energy, the multiple scattering 
contribution to the charge distribution is negligible. 
We find $\sigma_{{\mathrm{diff}}}$ = (16.3 $\pm$ 1.4)~$\mu$m, which agrees well with 
the diffusion length estimated from the doping in the epitaxial layer and the 
charge collection time~\cite{Battaglia:2008yt}.

\section{Measurement}

The detector charge-to-voltage conversion is 0.98~keV/ADC count or 26.7~$e^-$/ADC count 
at 6.25~MHz readout frequency, obtained by recording the position of the 5.9~keV 
full energy peak of a collimated 2.2~mCi $^{55}$Fe source.

\begin{figure}[ht!]
\begin{tabular}{c c}
\epsfig{file=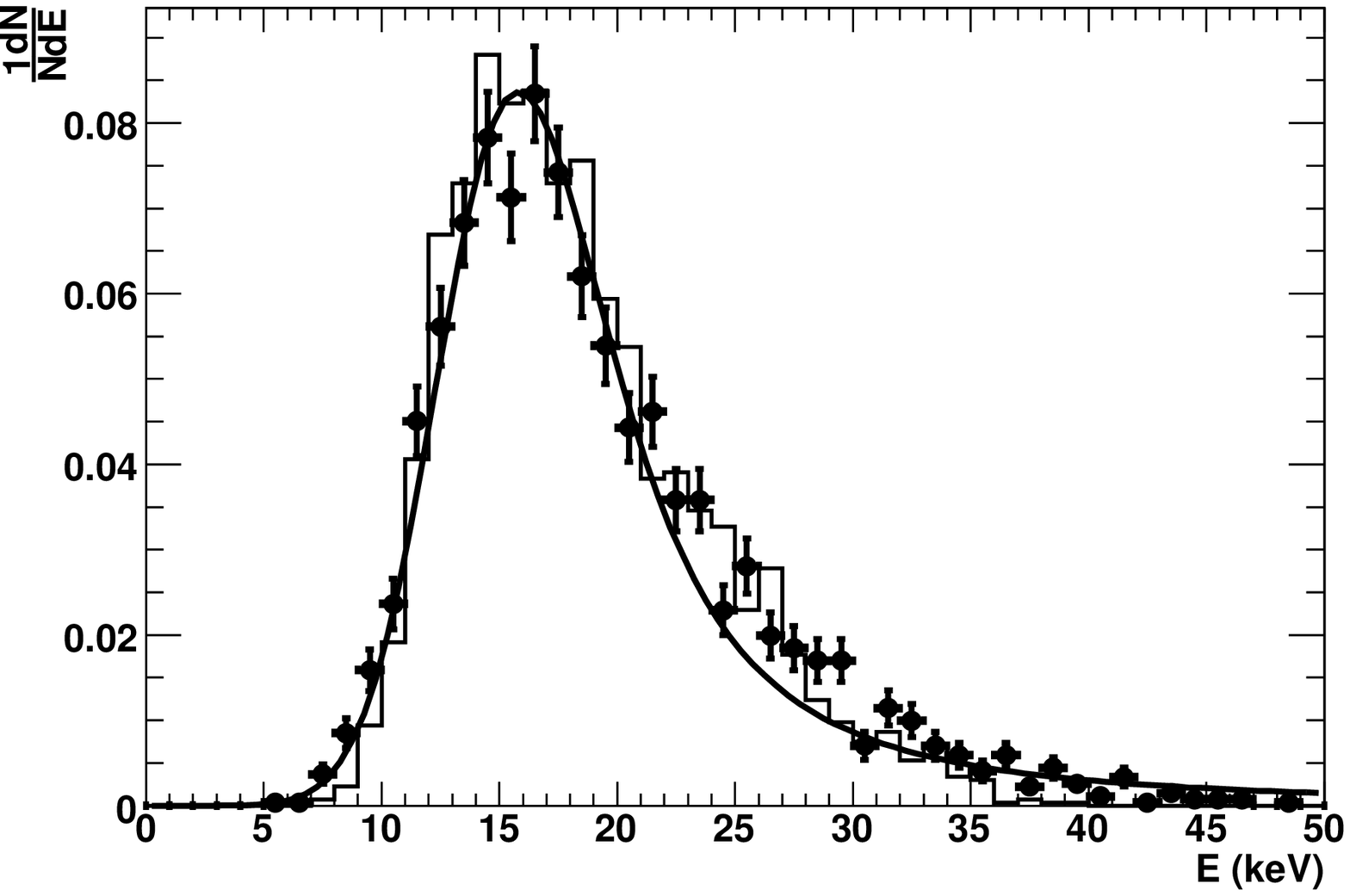,width=7.0cm} &
\epsfig{file=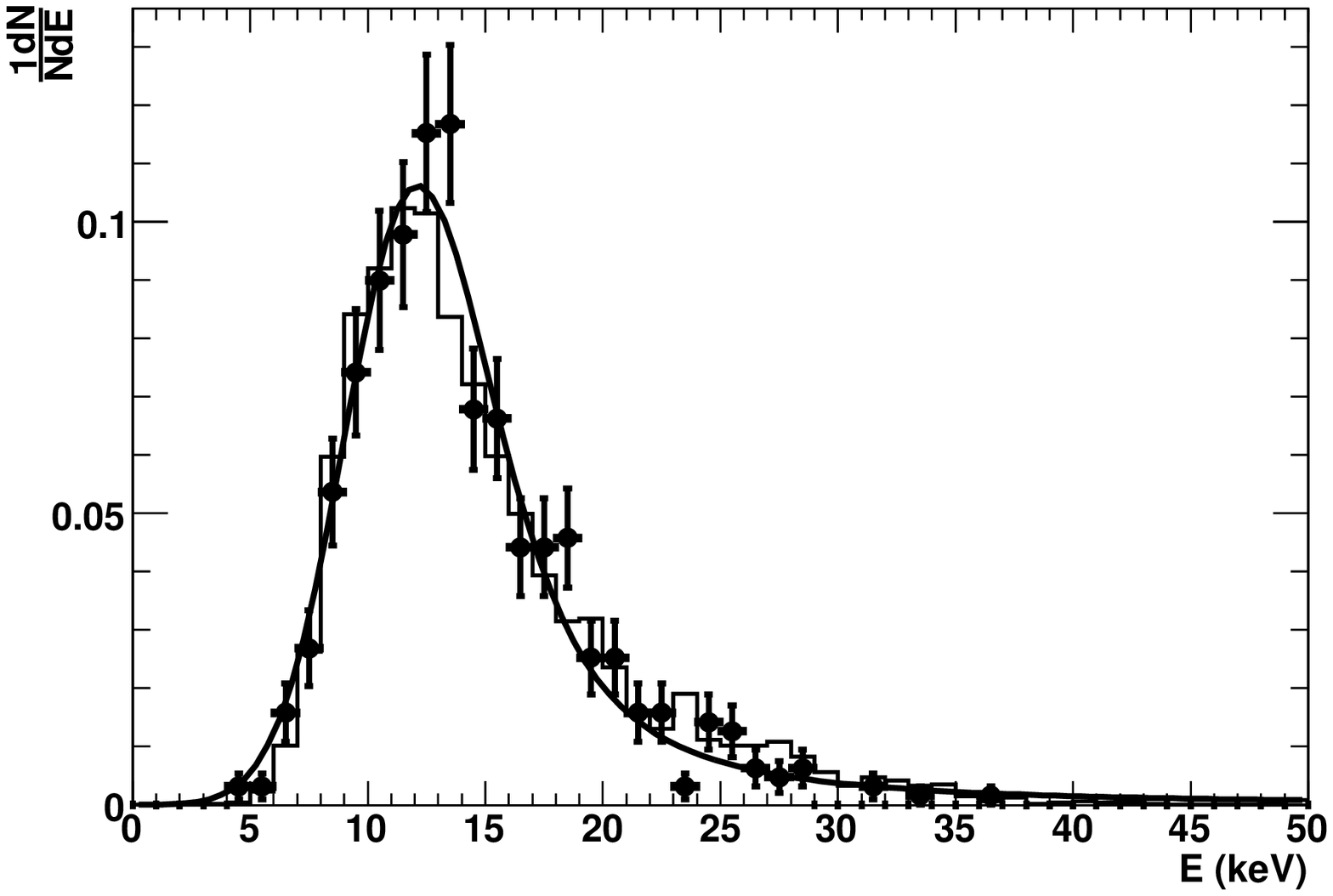,width=7.0cm} \\
\end{tabular}
\caption[]{Reconstructed deposited energy in a 3$\times$3 pixel 
matrix for 80~keV (left) and 100~keV electrons 
(right). The points with error bars show the data and the histogram
the result of the {\tt Geant-4} simulation. The continuous line shows
a Landau function convoluted with a Gaussian noise term fit to the data.}
\label{fig:landau}
\end{figure}

We use the TITAN test column at the National Center for Electron Microscopy (NCEM) 
to characterise the detector response to 80~keV and 100~keV electrons and validate
the simulation. The signal pulse height in a 3$\times$3 matrix around each seed 
pixel having a signal-to-noise in excess of 4.5 is shown in Figure~\ref{fig:landau} 
for data and simulation. The broadening of the energy distribution compared to electrons
of higher energy is evident, however the ratio of the Landau width to the Landau most 
probable value does not increase significantly compared to that for electrons of higher 
energy. We estimate the uncertainty on the number of electrons per pixel that can be 
reconstructed  from the measured 
pulse height in a single pixel. We simulate a flat field illumination by generating
multiple electrons hitting each pixel and reconstruct the pixel pulse height. This
accounts for cross-feed between neighbouring pixels due to charge diffusion and 
multiple scattering. We determine the number of electrons on each pixel by dividing 
the simulated pixel pulse height by the average pulse height induced by a single 
electron and study the distribution of the reconstructed number of electrons as a 
function of that simulated. We find that the relative uncertainties on this 
number scale from 0.17 (0.14) for 10 $e^-$/pixel to 0.12 (0.10) for 20 $e^-$/pixel
and to  0.08 (0.06) for 50 $e^-$/pixel at 80~keV (100~keV), respectively. These
results are comparable to relative uncertainties of 0.15, 0.11, 0.07 obtained for 
200~keV electrons. 

Finally, we determine the point spread function (PSF) following the same method 
discussed in~\cite{Battaglia:2008yt}. We reconstruct the image of a gold wire 
with a diameter measured to be (59.6$\pm$0.7)~$\mu$m and mounted parallel to the 
pixel columns, at a distance of $\simeq$~3~mm from the detector surface. 
The profile of the deposited energy in the pixels, measured across the wire allows us 
to determine the charge spread due to electron multiple scattering and charge carrier 
diffusion. 
\begin{table}[ht!]
\caption[]{Point spread function predicted by {\tt Geant 4} + {\tt PixelSim} 
and measured with data for 20~$\mu$m pixel pitch. The uncertainty quoted for 
simulation is the systematics from $\sigma_{{\mathrm{diff}}}$, that for data 
accounts for statistical and systematics from pixel response equalization.}
\begin{center}
\begin{tabular}{|c|c|c|}
\hline
Energy & {\tt Geant-4} + & Data     \\
(keV)  & {\tt PixelSim}  & 20~$\mu$m Pixels  \\ 
       & ($\mu$m)        & ($\mu$m) \\ \hline
~80    & 12.7 $\pm$ 0.5  & 12.1 $\pm$ 1.6    \\
100    & 13.2 $\pm$ 0.5  & 13.0 $\pm$ 1.7    \\
\hline
\end{tabular}
\end{center}
\label{tab:psfsim}
\end{table}
\begin{figure}[hb!]
\begin{center}
\epsfig{file=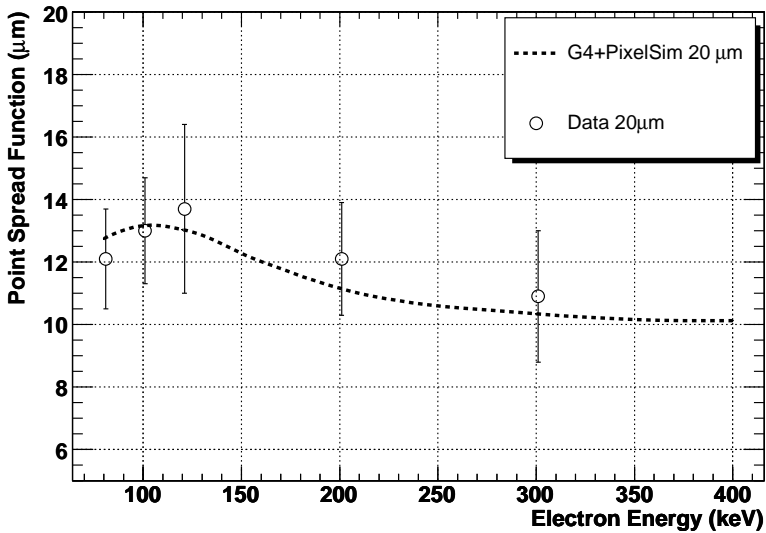,width=11.5cm}
\end{center}
\caption[]{Point spread function vs.\ electron energy with 
data (points with error bars) compared to simulation (lines) for 
20~$\mu$m pixels. Data points from 120~keV to 300~keV are from 
ref.~\cite{Battaglia:2008yt}. The new measurements at 80~keV and 
100~keV show the saturation of the point spread function contribution 
from multiple scattering in the sensor due to the reduced range of 
electrons as predicted by the simulation, shown by the line.}
\label{fig:psfit}
\end{figure}
We describe the measured pulse height on the pixel rows across the image projected 
by the wire with a box function having the same width as the measured wire diameter 
smeared by a Gaussian term, which describes the point spread function. The contrast 
factor, i.e.\ the ratio of maximum to minimum pulse height levels, for the pixels away 
from the wire shadow and for those exactly below the wire centre, respectively, are set 
to those observed in data and we perform a simple 1-parameter $\chi^2$ fit to extract 
the Gaussian width term, which gives the estimation of the PSF. 
Results are given in Table~\ref{tab:psfsim}. 
A good agreement is found between the measurement and the prediction 
from simulation. These results are compared with those obtained at higher 
energies, presented in~\cite{Battaglia:2008yt}. It is interesting to observe 
how the degradation of the PSF at decreasing energies, caused by multiple scattering, 
reaches a plateau around 120~keV. This is due to the decrease of the electron range 
with the particle energy, which limits the distance over which charge can be spread.
In fact, we measure a point spread function value for 80 and 100~keV electrons which 
is compatible with that measured at higher energies, as predicted by the simulation.
This result is quite encouraging for extending the application of CMOS pixel 
sensors to fast TEM imaging of organic and biological samples with low energy 
electrons.

\section*{Acknowledgements}

\vspace*{-0.1cm}

We wish to thank Thomas Duden, Rolf Erni and Zhongoon Lee.
This work was supported by the Director, Office of Science, of 
the U.S. Department of Energy under Contract No. DE-AC02-05CH11231.

\vspace*{-0.1cm}

 \nolinenumbers

\end{document}